\input{aipcheck}

\documentclass[
    ,final            % use final for the camera ready runs
  ]
  {aipproc}

\layoutstyle{8x11double}

\renewcommand\XFMtitleblock{%
  \XFMtitle
  \let\XFMoldpar\par
  \def\par{\XFMoldpar\def\par{\space
             (on behalf of the H.E.S.S.~Collaboration)\XFMoldpar}}%
   \XFMauthors
   \let\par\XFMoldpar
   \XFMaddresses
   \XFMabstract
   \vspace{5pt}%
   \XFMkeywords
   \XFMclassification
 }

\newcommand{\hess}{H.E.S.S.}

\newcommand{\xmm}{{\it XMM-Newton}}
\newcommand{\chandra}{{\it Chandra}}

\newcommand{\un}[1]{~\hspace{-1pt}\ensuremath{\mathrm{#1}}}

\newcommand{\xray}{X-ray}
\newcommand{\xrays}{X-rays}

\newcommand{\rxj}{RX~J1713.7-3946}
\newcommand{\velajr}{RX~J0852-4622}
\newcommand{\rcw}{RCW~86}
\newcommand{\snc}{G309.8-2.6}

\newcommand{\soua}{HESS~J1809-193}
\newcommand{\soub}{HESS~J1912+101}
\newcommand{\souc}{HESS~J1356-645}

\newcommand{\psra}{PSR~J1809-1917}
\newcommand{\psrb}{PSR~J1913+1011}
\newcommand{\psrc}{PSR~J1357-6429}

\newcommand{\be}{\begin{equation}}
\newcommand{\ee}{\end{equation}}
\newcommand{\ben}{\begin{eqnarray}}
\newcommand{\een}{\end{eqnarray}}
\newcommand{\bc}{\begin{center}}
\newcommand{\ec}{\end{center}}

\def\edot{$\dot{{\rm E}}$}

\def\d{$^\circ$}
\def\m{$^\prime$}
\def\s{$^{\prime\prime}$}
\def\hh{$^{\mathrm h}$}
\def\mm{$^{\mathrm m}$}
\def\ss{$^{\mathrm s}$}

\def\cm3{cm$^{-3}$~}

\def\eg{{\it e.g.~}}
\def\etal{et~al.~}
\def\ie{{\em i.e.~}}

\newcommand\aj{{AJ}}%
\newcommand\araa{{ARA\&A}}%
\newcommand\apj{{ApJ}}%
\newcommand\aap{{A\&A}}%
\newcommand\mnras{{MNRAS}}%
\begin{document}

\title{Pulsar Wind Nebula candidates \\
       recently discovered by H.E.S.S.}



\author{M. Renaud}{
  address={Max-Planck-Institut f\"ur Kernphysik, Postfach 103980,
  69029 Heidelberg, Germany}, email={mrenaud@mpi-hd.mpg.de}}

\author{S. Hoppe}{
  address={Max-Planck-Institut f\"ur Kernphysik, Postfach 103980,
  69029 Heidelberg, Germany}}

\author{N. Komin}{
  address={CEA/DSM/IRFU, CE Saclay, F-91191 Gif-sur-Yvette,
Cedex, France}}

\author{E. Moulin}{
  address={CEA/DSM/IRFU, CE Saclay, F-91191 Gif-sur-Yvette,
Cedex, France}}

\author{V. Marandon}{
  address={APC-UMR 7164, 10, rue Alice Domon et Leonie Duquet,
  F-75205 Paris Cedex 13}}

\author{and A.-C. Clapson}{
  address={Max-Planck-Institut f\"ur Kernphysik, Postfach 103980,
  69029 Heidelberg, Germany}}

\begin{abstract}

\hess~is currently the most sensitive instrument in the
very-high-energy (VHE; E $>$ 100 GeV) gamma-ray domain and has
revealed many new sources along the Galactic Plane, a significant
fraction of which seems to be associated with energetic pulsars.
HESS~J1825-137 and Vela~X are considered to be the prototypes of
such sources in which the large VHE nebula results from the whole
history of the pulsar wind and the supernova remnant host, both
evolving in a complex interstellar medium. These nebulae are seen
to be offset from the pulsar position and, for HESS~J1825-137, a
spectral steepening at increasing distance from the pulsar has
been measured. In this context, updated \hess~results on two
previously published sources, namely \soua~and \soub, and
preliminary results on the newly discovered \souc, are presented.
These extended VHE sources are thought to be associated with the
energetic pulsars \psra, \psrb~and \psrc, respectively. Properties
of each source in the VHE regime, together with those measured in
the radio and \xray~domains, are discussed.

\end{abstract}

\maketitle

%%%%%%%%%%%%%%%%%%%%%%%%%%%%%%%%%%%%%%%%%%%%
%% MAINMATTER
%%%%%%%%%%%%%%%%%%%%%%%%%%%%%%%%%%%%%%%%%%%%

\section{Introduction}
\label{s:1}

The recently opened astronomical window at VHE energies, with the
help of the latest generation of Imaging Atmospheric Cherenkov
Telescopes (IACTs) such as \hess~(High Energy Stereoscopic
System), has led to the discovery of more than 70 sources
\cite{c:hinton08}. Among the about 50 Galactic objects
\cite{c:chaves08} and besides the well-known shell-type supernova
remnants (SNRs) \rxj, \velajr~and \rcw, a significant fraction
seems to be associated with energetic pulsars. These objects can
generate bubbles of relativistic particles and magnetic field when
their ultra-relativistic wind interacts with the surrounding
medium (SNR or interstellar medium, hereafter ISM) (see
\cite{c:gaensler06,c:bucciantini08} for recent reviews). Their
confinement leads to the formation of strong shocks, which can
accelerate particles up to hundreds of TeV and beyond, thus
generating luminous nebulae seen across the entire electromagnetic
spectrum: in the synchrotron emission from radio to hard \xrays,
and through the inverse Compton process and potentially $\pi^{0}$
decay from p-p interactions \citep{c:horns06}, in the VHE domain.

On one hand, recent advances in the study of pulsar wind nebulae
(PWNe) have been made with the wealth of radio and
\xray~observations, revealing the complex morphology of these
sources at the arcsecond scale \cite{c:gaensler06}. On the other
hand, complementary VHE observations of these PWNe allow the
spectrum of accelerated particles to be probed and to investigate
the associated magnetic field distribution (see \citep{c:dd08} for
a recent discussion in this regard). Since VHE-emitting electrons
are usually less energetic than \xray-emitting ones, they do not
suffer from severe radiative losses and the majority of them may
survive from (and hence probe) early epochs of the PWN evolution.
The discovery of VHE gamma-rays from HESS~J1825-137 \cite{c:j1825}
and Vela~X \cite{c:velax}, both offset from their respective
pulsars, confirm predictions
\cite{c:blondin01,c:gaensler03,c:vds04} that anisotropic reverse
shocks and/or high pulsar velocity can lead to the formation of a
relic PWN, presumably emitting in the VHE domain.

In this contribution, updated \hess~results on two published
sources, namely \soua~and \soub, are presented. Preliminary
results on the newly discovered \souc~are also shown. Standard
quality selection critera (see \eg~\cite{c:gps06}) were applied
for each data set, background modelling was performed according to
the methods discussed in \cite{c:berge07}, and imaging and
spectral results on each source were cross-checked through two
independent data analyses, namely the Hillas \cite{c:hofmann99}
and Model 2D \cite{c:denaurois03} methods. These three sources are
thought to be associated with pulsars \psra, \psrb~and \psrc,
respectively, and their properties as offset VHE PWN
\textit{candidates} are then discussed.

\section{HESS~J1809-193}
\label{s:2}

\begin{figure}[!htb]

  \includegraphics[height=.24\textheight]{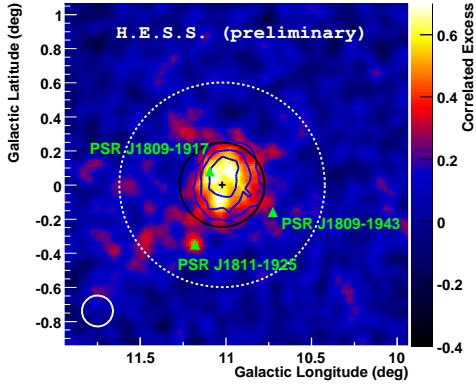}

  \caption{Image of the VHE $\gamma$-ray excess in Galactic coordinates,
           centered on \soua~and smoothed with a Gaussian of width 0.06\d.
           The blue contours correspond to the 4, 6 and 8 $\sigma$ levels
           calculated for an integration radius $\theta_{{\rm cut}}$ of 0.1\d.
           The black cross indicates the best fit position of the source
           centroid together with its statistical errors. The intrinsic source
           rms is shown by the black solid circle after fitting with a symmetric
           Gaussian. The dotted circle indicates the region over which
           the spectrum shown in Figure \ref{f:j1809_2} has been extracted. The
           white solid circle illustrates the 68\% containment radius of the
           point spread function resulting from the convolution with the
           Gaussian profile used for smoothing the image. Due to the presence
           of some large-scale emission, the size of this region differs from
           that initially published. This explains the difference in the spectral
           normalization reported in Table \ref{t:charac}.}
  \label{f:j1809_1}

\end{figure}

\begin{figure}[!htb]

  \includegraphics[height=.255\textheight]{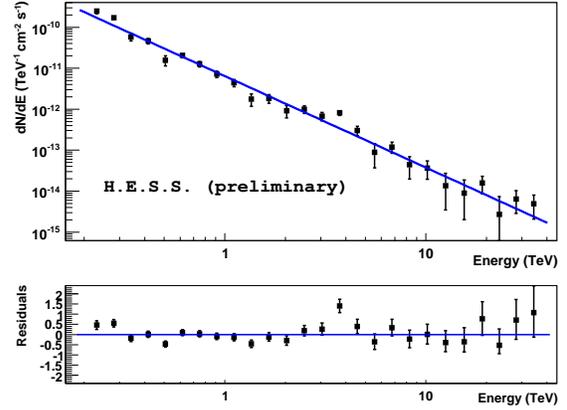}

  \caption{Differential energy spectrum of \soua. The data points
           were fitted with a power law whose best fit is shown
           with the blue solid line. Also shown are the residuals
           in the bottom panel.}
  \label{f:j1809_2}

\end{figure}

\soua~was initially discovered during the systematic search for
VHE emission from pulsars in the Galactic Plane Survey performed
with \hess~\cite{c:j1809}. \psra, with a characteristic age
$\tau_{c}$ of 51 kyr and a spin-down power \edot~of 1.8 $\times$
10$^{36}$ erg s$^{-1}$, was considered as the most likely
counterpart. Since the original discovery, new data were taken and
the total livetime now amounts to 41 hours (versus 25 hours
before). Figures \ref{f:j1809_1} and \ref{f:j1809_2} show the
updated \hess~image of the smoothed excess counts centered on
\soua~and its spectrum measured between 300\un{GeV} and
30\un{TeV}, respectively. General information on the source, in
relation to \psra, is summarized in Table \ref{t:charac}.

As discussed in \cite{c:j1809}, the region around \soua~contains
several candidates for TeV emission. Besides \psra, SNR candidates
recently revealed in radio \cite{c:brogan06,c:helfand06} lie
within the VHE source extent. These SNRs may contribute to the VHE
emission observed by \hess. However, the discovery of large-scale
($\sim$6\m) \xray~emission, surrounding the compact PWN associated
with \psra, and extending toward the VHE centroid
\cite{c:komin07,c:kp07}, strengthens the scenario of a relic PWN
crushed due to an inhomogeneous SNR interior \cite{c:blondin01}.
Then, \soua~seems to share several similarities with firmly
established VHE PWNe such as HESS~J1825-137 \cite{c:j1825}. Deeper
radio observations could detect the synchrotron emission from this
potential relic nebula.

\section{HESS~J1912+101}
\label{s:3}

\begin{figure}[!htb]

  \includegraphics[height=.22\textheight]{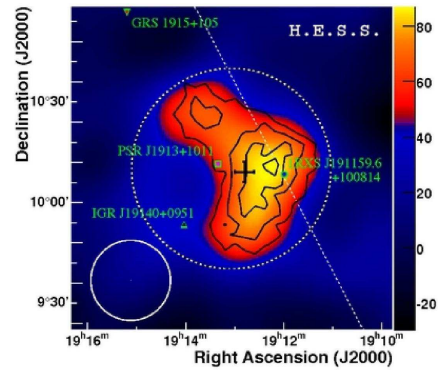}

  \caption{Image of the VHE $\gamma$-ray excess centered on \soub,
           smoothed with a Gaussian of width 0.13\d. The blue contours
           correspond to the 3.5, 4.5, 5.5 and 6.5 $\sigma$ levels calculated
           for $\theta_{{\rm cut}}$ = 0.22\d. The black cross
           indicates the best fit position of the source centroid
           together with its statistical errors. The dotted circle
           indicates the region over which the spectrum shown in Figure
           \ref{f:j1912_2} has been extracted. The white solid circle
           has the same meaning as in Figure \ref{f:j1809_1}.}
  \label{f:j1912_1}

\end{figure}

\begin{figure}[!htb]

  \includegraphics[height=.25\textheight]{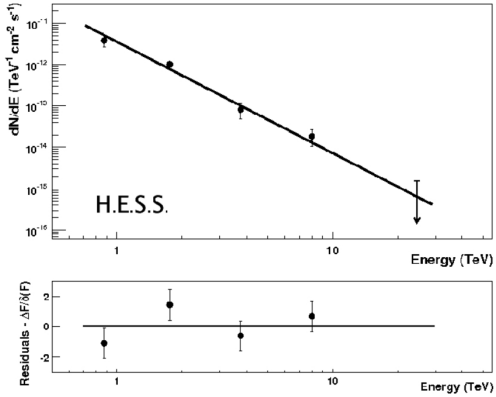}

  \caption{Differential energy spectrum of \soub. The data points
           were fitted with a power law whose best fit is shown
           with the black solid line. Also shown are the residuals
           in the bottom panel.}
  \label{f:j1912_2}

\end{figure}

\soub~was discovered during the continuation of the \hess~Galactic
Plane Survey \cite{c:j1912}. The VHE source, with a post-trials
significance of 5.5 $\sigma$ for a livetime of 21 hours, is fairly
extended (intrinsic Gaussian width of 0.26 $\pm$ 0.03\d), as shown
in Figure \ref{f:j1912_1}. Its spectrum (Figure \ref{f:j1912_2})
is well fitted with a power law with spectral index of 2.7 $\pm$
0.2.

\psrb~($\tau_{c}$ = 170 kyr and \edot~= 2.9 $\times$ 10$^{36}$ erg
s$^{-1}$), slightly offset (at the $\sim$ 3 $\sigma$ confidence
level) from the centroid of the VHE emission, is thought to be the
most likely counterpart, both in positional and energetic
connection to \soub~(see Table \ref{t:charac}). If true, this
source would be the oldest candidate for a VHE emitting PWN.
Energetic pulsars usually exhibit a more compact \xray~nebula, but
such a PWN has not yet been found with the existing
\xray~observations. A pulsar kick velocity of $\sim$ 60 km
s$^{-1}$ would explain the observed offset between \psrb~and the
VHE centroid. On the other hand, as discussed in \cite{c:j1912},
the hypothesis of an inhomogeneous surrounding medium seems
plausible given the clumpy molecular material revealed by
$^{13}$CO line emission (see Figure 1 in \cite{c:j1912}) at a
similar distance as \psrb.

\section{HESS~J1356-645}
\label{s:4}

\begin{figure}[!htb]

  \includegraphics[height=.24\textheight]{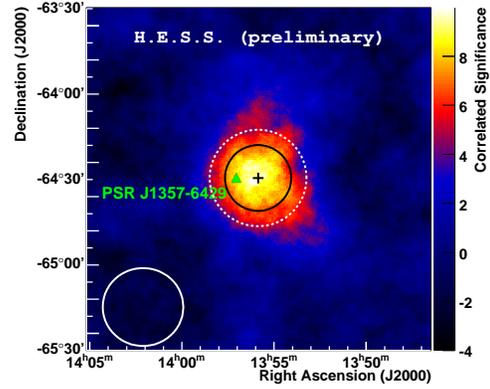}

  \caption{Image of VHE correlated (pre-trials) significance ($\theta_{{\rm cut}}$
           = 0.22\d) centered on \souc. The black cross indicates the
           best fit position of the source centroid together with its
           statistical errors. The intrinsic (rms) Gaussian width of the
           source is shown by the black solid circle. The dotted circle
           indicates the region over which the spectrum shown in Figure
           \ref{f:j1356_2} has been extracted. The white solid circle
           has the same meaning as in Figure \ref{f:j1809_1}. The position
           of \psrc~is marked in green.}
  \label{f:j1356_1}

\end{figure}

\begin{figure}[!htb]

  \includegraphics[height=.24\textheight]{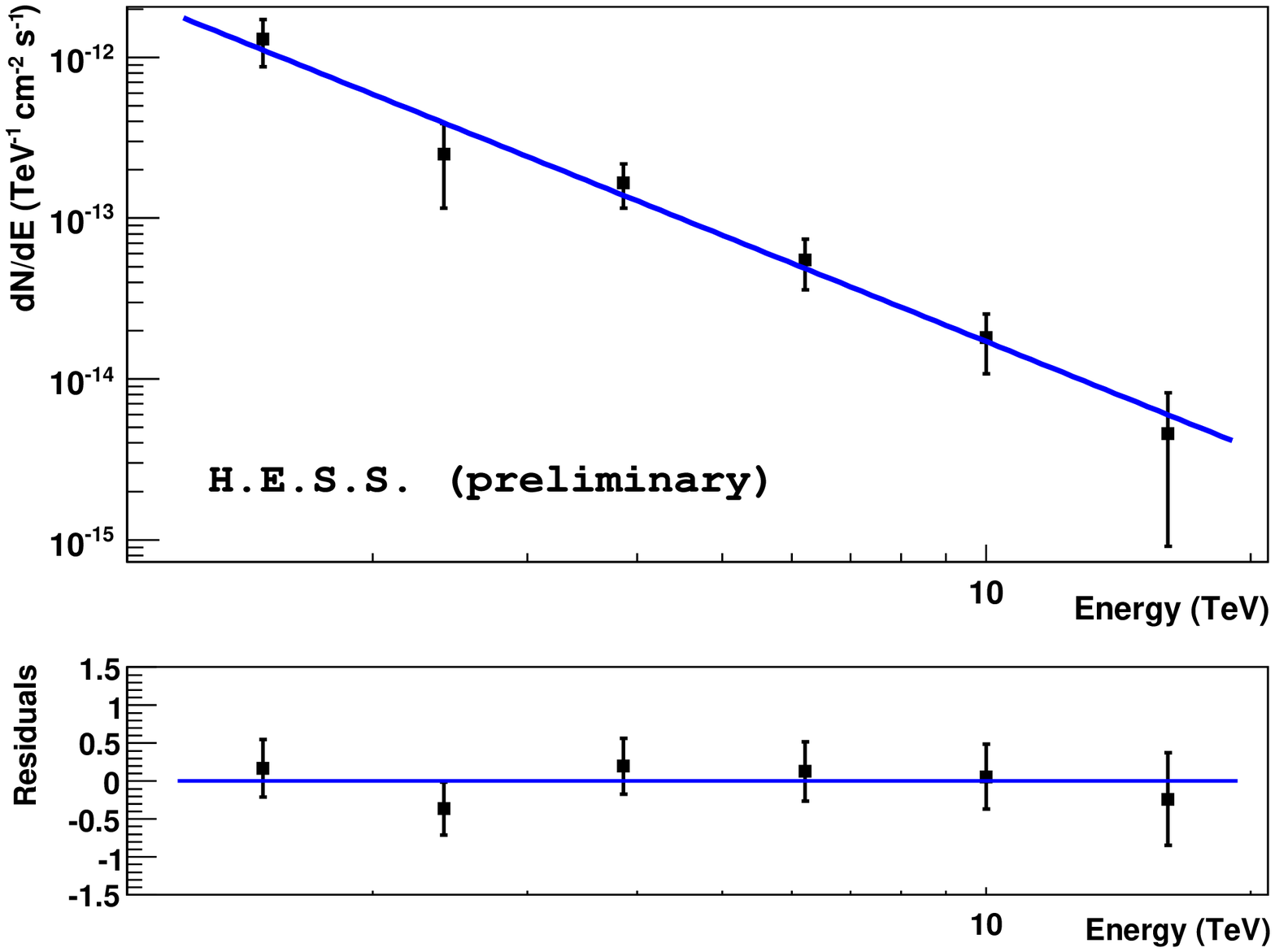}

  \caption{Differential energy spectrum of \souc. The data points
           were fitted with a power law whose best fit is shown
           with the blue solid line. Also shown are the residuals
           in the bottom panel.}
  \label{f:j1356_2}

\end{figure}

\begin{table}[!htb]
\caption{Characteristics of the three \hess~sources discussed
here. Positions (with statistical errors of $\sim$ 2\m~for each
axis) and intrinsic source rms were obtained after fitting with a
2D symmetric Gaussian. The offset is between the centroid of the
VHE emission and the associated pulsar. L$_{{\rm 1-10
TeV}}$/\edot~denotes the efficiency as the ratio of the
$\gamma$-ray luminosity of the VHE source to the pulsar spin-down
power. Properties of the three pulsars were taken from the online
ATNF Pulsar Catalogue (version 1.33). Distance estimates given
below (in units of kpc) were calculated according to the Galactic
electron distribution model of Cordes \& Lazio \cite{c:cl02}. The
uncertainties shown in parentheses are statistical errors only.
Systematic errors on spectral indexes are typically 0.2.}
\label{t:charac}

\centering
\begin{tabular}{|c|c|c|c|c|c|c|c|}
\hline

H.E.S.S. &   R.A.  &   Dec.  & Source rms & Offset &               Norm(1 TeV)                    & Slope & L$_{{\rm 1-10 TeV}}$/\edot \\
source   & (J2000) & (J2000) &   (deg.)   & (deg.) & \footnotesize{(cm$^{-2}$s$^{-1}$TeV$^{-1}$)} &       &         (\%)               \\
\hline

J1809-193 & 18\hh09\mm52\ss & -19\d23\m42\s & 0.25(0.02) & 0.11(0.02) &  6.4(3.0)$\times$10$^{-12}$ & 2.23(0.05) & 1.5(0.8)$\times$$d^{2}_{3.5}$ \\
J1912+101 & 19\hh12\mm49\ss &  10\d09\m06\s & 0.26(0.03) & 0.15(0.05) &  3.5(0.6)$\times$10$^{-12}$ &  2.7(0.2)  & 0.5(0.2)$\times$$d^{2}_{4.5}$\\
J1356-645 & 13\hh56\mm00\ss & -64\d30\m00\s & 0.20(0.02) & 0.11(0.03) &  2.7(0.9)$\times$10$^{-12}$ &  2.2(0.2)  & 0.2(0.1)$\times$$d^{2}_{2.5}$ \\

\hline
\end{tabular}
\end{table}

Preliminary results on the newly discovered source \souc~are here
presented. The livetime of the observations amounts to 10 hours,
which leads to the detection of a $\sim$ 8.5 $\sigma$
(post-trials, see \cite{c:gps06}) extended source, as shown in
Figure \ref{f:j1356_1}. A power law fit of its spectrum (see
Figure \ref{f:j1356_2}) gives a spectral index of 2.2 $\pm$ 0.2
and an integral flux between 1 and 10\un{TeV} of about 11\% of
that of the Crab Nebula in the same energy band.

Since \souc~is located at $\sim$ 2.5\d~below the Galactic Plane,
the identification of counterparts is easier than for the two
previous sources. The only potential counterpart known is the
recently discovered young ($\tau_{c}$ = 7.3 kyr) and energetic
(\edot~= 3.1 $\times$ 10$^{36}$ erg s$^{-1}$) 166\un{ms} pulsar
\psrc~\cite{c:camilo04}, lying only $\sim$ 0.11\d~from the VHE
centroid. General results from \chandra~and \xmm~observations were
reported in \cite{c:esposito07,c:zavlin07}. Marginal evidence of
diffuse \xray~emission around the pulsar was found
\cite{c:zavlin07}. \souc~is coincident with extended radio
emission at 2.4\un{GHz}, originally catalogued as an SNR candidate
(\snc, \cite{c:duncan97}). The inspection of archival public radio
images from the Molonglo Galactic Plane Survey at 843\un{MHz}
(MGPS-2, \cite{c:murphy07}) and from the Parkes-MIT-NRAO (PMN)
survey at 4.85\un{GHz} \cite{c:griffith93} has revealed an
extended structure which, although faint, appears coincident with
the observed VHE emission as well. A study of the radio spectral
index from these three images is on-going and could help to
constraint the nature of \souc.

\section{Discussion}
\label{s:5}

The main characteristics of the three \hess~sources, in relation
with their respective pulsars, are summarized in Table
\ref{t:charac}. The most uncertain pulsar-VHE association is
certainly that between \psrb~and \soub, since no \xray~emission
has been found neither from the pulsar itself nor from any
putative PWN. For the two others, indications of such an
association exist, through the extended \xray~emission towards
\soua, and through the extended radio structure coincident with
\souc. The VHE efficiencies of these three \hess~sources are of
the same order as that measured in other VHE PWNe, \ie around one
percent. This qualitative argument shows that these pulsars are
energetic enough to power the observed VHE extended sources,
thought to be the relic PWNe probing their past evolution. Deeper
follow-up observations, mainly in radio and \xrays, are necessary
in order to reveal the nature of these sources. As discussed in
\cite{c:hinton08}, PWNe have now emerged as the largest population
of galactic VHE sources. The three \hess~sources presented here
will certainly benefit population studies
\cite{c:grenier08,c:mattana08}, which may provide new insights on
these complex sources.

\begin{theacknowledgments}

The support of the Namibian authorities and of the University of
Namibia in facilitating the construction and operation of \hess~is
gratefully acknowledged, as is the support by the German Ministry
for Education and Research (BMBF), the Max Planck Society, the
French Ministry for Research, the CNRS-IN2P3 and the Astroparticle
Interdisciplinary Programme of the CNRS, the U.K. Science and
Technology Facilities Council (STFC), the IPNP of the Charles
University, the Polish Ministry of Science and Higher Education,
the South African Department of Science and Technology and
National Research Foundation, and by the University of Namibia. We
appreciate the excellent work of the technical support staff in
Berlin, Durham, Hamburg, Heidelberg, Palaiseau, Paris, Saclay, and
in Namibia in the construction and  operation of the equipment.

\end{theacknowledgments}

\bibliographystyle{aipproc}

\begin{thebibliography}{9}

\bibitem{c:hinton08} Hinton, J. 2008, New Journal of Physics, in press
\bibitem{c:chaves08} Chaves, R.C.G., \etal (\hess~Collaboration) 2008, these proceedings
\bibitem{c:gaensler06} Gaensler, B.M., \& Slane, P.O. 2006, \araa, 44, 17
\bibitem{c:bucciantini08} Bucciantini, N. 2008, AIP Conference Proceedings, Vol. 983, pp. 186-194
\bibitem{c:horns06} Horns, D., \etal 2006, \aap, 451, L51
\bibitem{c:dd08} de Jager, O.C., \& Djannati-Ata\"i, A. 2008, Springer Lecture Notes on
                 "Neutron Stars and Pulsars: 40 years after their discovery", ed. W.
                 Becker, in press
\bibitem{c:j1825} Aharonian, F., \etal 2006, \aap, 460, 365
\bibitem{c:velax} Aharonian, F., \etal 2006, \aap, 448, L43
\bibitem{c:blondin01} Blondin, J.M., Chevalier, R.A., \& Frierson, D.M. 2001, \apj, 563, 806
\bibitem{c:gaensler03} Gaensler, B.M., \etal 2003, \apj, 588, 441
\bibitem{c:vds04} van der Swaluw, E., Downes, T.P., \& Keegan, R. 2004, \aap, 420, 937
\bibitem{c:gps06} Aharonian, F., \etal 2006, \apj, 636, 777
\bibitem{c:berge07} Berge, D., \etal 2007, \aap, 466, 1219
\bibitem{c:hofmann99} Hofmann, W., \etal 1999, Astropart. Phys., 12, 135
\bibitem{c:denaurois03} de Naurois, M., \etal 2003, 28th ICRC proceeding p.2907
\bibitem{c:j1809} Aharonian, F., \etal 2007, \aap, 472, 489
\bibitem{c:brogan06} Brogan, C.L., \etal 2006, \apj, 639, L25
\bibitem{c:helfand06} Helfand, D.J., \etal 2006, \aj, 131, 2525
\bibitem{c:komin07} Komin, N., \etal (\hess~Collaboration) 2007, 30th ICRC proceeding, in press
\bibitem{c:kp07} Kargaltsev, O., \& Pavlov, G.G. 2007, \apj, 670, 655
\bibitem{c:j1912} Aharonian, F., \etal 2008, \aap, 484, 435
\bibitem{c:camilo04} Camilo, F., \etal 2004, \apj, 611, L25
\bibitem{c:esposito07} Esposito, P. \etal 2007, \aap, 467, L45
\bibitem{c:zavlin07} Zavlin, V.E. 2007, \apj, 665, L143
\bibitem{c:duncan97} Duncan, A.R. \etal 1997, \mnras, 287, 722
\bibitem{c:murphy07} Murphy, T. \etal 2007, \mnras, 382, 382
\bibitem{c:griffith93} Griffith, M.R., \& Wright, A.E. 1993, \aj, 105, 1666
\bibitem{c:cl02} Cordes, J.M., \& Lazio, T.J. 2002, astro-ph/0207156
\bibitem{c:grenier08} Grenier, I.A. 2008, 30th ICRC proceeding, in press
\bibitem{c:mattana08} Mattana, F., \etal 2008, these proceedings

\end{thebibliography}

\end{document}